\documentclass[aps,prl,showpacs,twocolumn,groupedaddress]{revtex4}

\usepackage{graphicx}
\usepackage{dcolumn}
\usepackage{bm}

\begin{document}


\title{New Magnetic Excitations in the Spin-Density-Wave of Chromium}


\author{H.~Hiraka$^1$, P.~B\"oni$^2$, Y.~Endoh$^1$, M.~Fujita$^3$, K.~Yamada$^3$,
        and G.~Shirane$^4$}
\affiliation{ $^1$Institute for Materials Research, Tohoku University,
Katahira, Aoba-ku, Sendai 980-8577, Japan\\
   $^2$Physik-Department E21, Technische Universit\"at M\"unchen, D-85747 Garching, Germany \\
   $^3$Institute for Chemical Research, Kyoto University, Gokashou, Uji 611-0011, Japan\\
   $^4$Physics Department, Brookhaven National Laboratory, Upton, NY 11973, USA
}


\date{\today}

\begin{abstract}

Low-energy magnetic excitations of chromium have been
reinvestigated with a single-{\boldmath $\it Q$} crystal using neutron scattering technique.
In the transverse spin-density-wave phase
a new type of well-defined magnetic excitation is
found around ($0,0,1$) with a weak dispersion
perpendicular to the wavevector of the incommensurate structure.
The magnetic excitation has an energy gap
of $\omega \simeq 4$~meV and at ($0,0,1$) exactly corresponds to the Fincher mode
previously studied only along the incommensurate wavevector.

\end{abstract}


\maketitle



\begin{figure}[h]
\centering
\includegraphics[scale=0.42,clip]{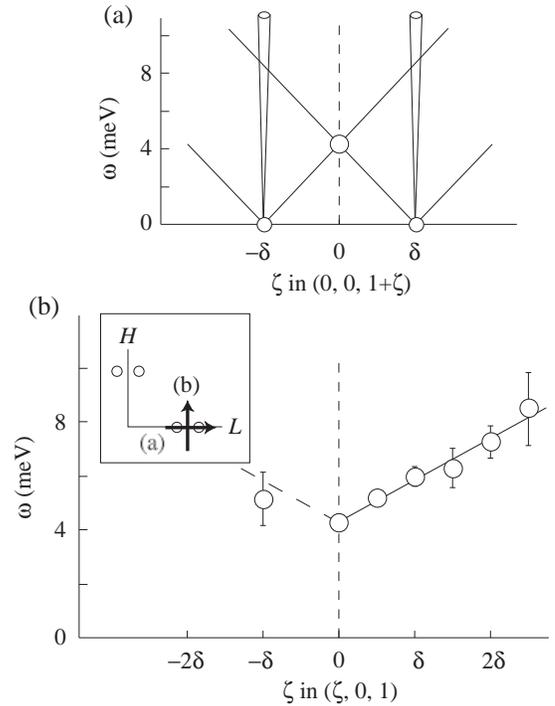} 
\caption{Energy dependence of magnetic excitations in the single-$\bf Q$
TSDW phase of Cr.
(a) The cones at the incommensurate positions ($0,0,1\pm\delta$) indicate
the high-velocity spin excitations. 
The solid lines represent
the proposed dispersion of the FB mode, which
cross at the commensurate position ($0,0,1$) at $\simeq$ 4 meV.
(b) The data points for the measurements transverse to $\bf Q_\pm$ 
indicate excitations with a gapped dispersion. 
The inset defines the scans in reciprocal space.
Most measurements have been performed around the ($0,0,1$) Bragg point. 
} \label{Fig1}
\end{figure}

The spin-density-wave (SDW) in Cr is one of the most fascinating
subjects in condensed matter physics. It has a history of  long
and continuing research.~\cite{fawcett88} 
In spite of the simple
body-centered cubic structure with a  lattice constant $a = 2.88$~\AA,
Cr and its alloys show interesting magnetic behaviors.~\cite{Werner67,fawcett91}
Below the Neel temperature {\it T}$_{\rm N} = 311$~K,
an incommensurate antiferromagnetic
structure develops due to a transverse spin-density-wave (TSDW) with the
moments oriented perpendicular to the ordering wavevector
${\bf Q}_{\pm} = (2\pi/a)(0, 0, 1\pm\delta)$
($\delta\simeq 0.048$ at $T = 100$~K) (see Fig.~\ref{Fig1}(a)).
At $T_{\rm sf} = 121$~K a spin-flop transition takes
place to a longitudinal spin-density-wave (LSDW) phase with the moments
along ${\bf Q}_{\pm}$.

The magnetic cross section in this system also shows a
surprisingly rich behavior. The magnetic excitations from these
SDW ordered states emerge from the incommensurate positions with
high mode-velocities (Fig.~\ref{Fig1}(a), two cones). In the TSDW phase this
metallic antiferromagnet exhibits two types of magnetic fluctuations with the 
polarization transverse and longitudinal relative
to the spin direction. 
Recently, using polarized neutron
scattering in the TSDW phase, it was confirmed that the velocity
of the transverse-mode excitations is significantly higher than
the velocity of the longitudinal excitations.~\cite{boeni98}.

In addition to the incommensurate scattering with large energy scale,
Fincher {\it et al}. observed a resonance like scattering
localized at the commensurate position
($0, 0, 1$) and at $\omega = 4$~meV, (Fig.~\ref{Fig1}(a), open circle)
in the TSDW phase.~\cite{fincher79,fincher81}
Later on Burke {\it et al}. reinvestigated the low energy excitations and
concluded that the Fincher-excitation at ($0,0,1$) was part of dispersion
curves for magnetic modes that emanate symmetrically from the
($0,0,1\pm\delta$) positions at the incommensurate wavevectors
(Fig.~\ref{Fig1}(a), lines).~\cite{burke83}
Although many neutron scattering experiments have been performed
around ($0,0,1$)~\cite{pynn92,sternlieb93,lorenzo94,sternlieb95,fukuda96,boeni98} 
and many interesting results were presented, 
no simple and conclusive explanation was obtained for the origin and details of
the Fincher-Burke(FB) mode.
Among them the most recent polarized beam and high-resolution
measurements provide several constraints for the FB-mode: 
i) the
mode has longitudinal polarization thus ruling out the possibility
of magneto-vibrational scattering,~\cite{pynn92,boeni98} 
ii) the symmetric
branches for $Q < Q_-$ and $Q > Q_+$ do not exist,~\cite{sternlieb93} and 
iii)
the dispersion below 4 meV is absent, i.e. the FB-mode has an
energy gap of 4 meV.~\cite{boeni02}
However, there still exists substantial disagreement between different
experiments, which precludes a full understanding of the magnetic
excitations of Cr.

Similar to the experimental side
there exist contrasting discussions between theories even in the ground state.
For example, although the incommensurate
ordering can be explained by the nesting properties of the electron and
hole Fermi-surfaces~\cite{holroyd80},
a recent density-functional investigation
predicts a commensurate structure.~\cite{hafner}
Concerning the variety of magnetic excitations
Fishman and Liu~\cite{fishman96} succeeded in calculating
the incommensurate excitations and assigning the longitudinal modes as
being phason modes.
In addition, they predicted a large number of possible interband transitions. 
However,
since the accuracy of present-day band-calculations does not allow to
calculate the low energy spectrum in Cr with high-precision,
there is still no acceptable model to explain the FB-excitations.

In this paper, we report a new type of magnetic excitations in the SDW state of Cr.
The low energy magnetic excitations of Cr were explored 
using a large single crystal with a single-{\bf Q} structure.  
The original main target was to study the magnetic cross section 
at the so-called "silent" satellites first investigated 
by Sternlieb {\it et al}.~\cite{sternlieb95}
We, therefore, have taken the data near the incommensurate positions ($\pm\delta,0,1$).  
Quite surprisingly, the well-defined gapped energy spectrum at ($\pm\delta,0,1$) was 
also observed with a weak dispersion with $\bf Q$ 
towards the Fincher excitation at ($0,0,1$) as shown in Fig.~\ref{Fig1}(b). 
Therefore, the Fincher mode has a clear dispersion 
perpendicular to the incommensurate wavevectors. 
This new observation of a gaped excitation provides new
restrictions on the origin of the FB-mode and definitely buries
all previous interpretations.

%
\begin{figure}[h]
\centering
 \includegraphics*[scale=0.5,clip]{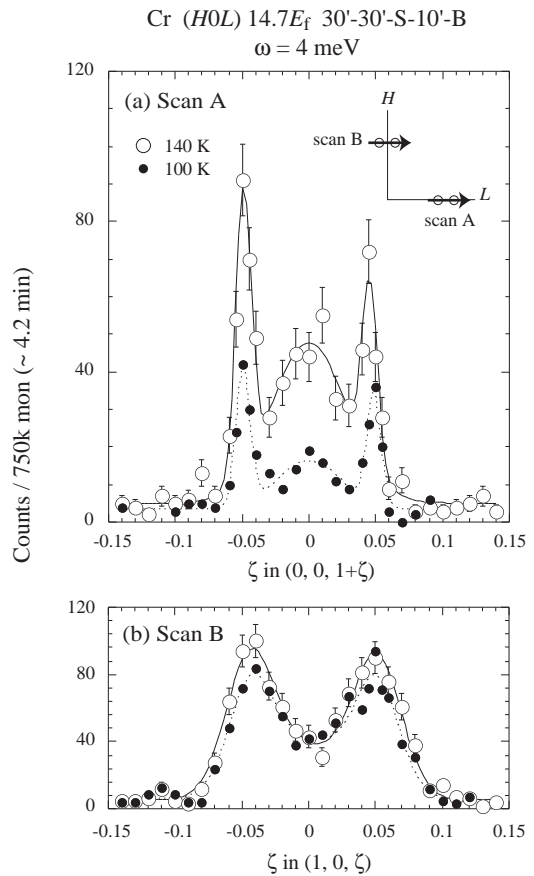}
\caption{
The scans for constant $\omega = 4$ meV 
with scattering vectors $\bf Q$ (a) along (scan A in the inset)
 and (b) perpendicular (scan B in the inset)
to the incommensurate ordering vector $\bf Q_\pm$
in the TSDW ($T = 140$~K) and LSDW ($T = 100$~K) phases.
The Fincher-mode is observed by the scan A only in the TSDW phase 
(open circles in (a)). 
The weak commensurate scattering intensity in the 
LSDW phase of (a) does not signify any excitations
(see text). (Ref. Fincher{\it et al}.~\cite{fincher81}) 
}
\label{Fig2}
\end{figure}

The inelastic neutron scattering was performed using a large cylindrical single
crystal of Cr from Johnson-Matthey Co. with a diameter of 10 mm,
a length of 50~mm along [$0,1,0$] direction
($V \simeq 4$ cm$^3$) and a mosaic $\eta
\simeq 40'$ on the triple-axis spectrometer TOPAN at JRR-3M in Tokai, Japan. 
In order to produce a single-$\bf Q$ sample the crystal was
cooled through ${\it T}_{\rm N}$ in a field of 14~T. 
The field work was kindly accomplished in
cooperation with the High Field Laboratory for Superconducting Materials, Tohoku University.
The crystal was aligned with the [$1,0,0$] and the [$0,0,1$] crystallographic directions in
the scattering plane. 
The [$0,0,1$] direction was selected to be parallel to $\bf Q_\pm$. 
Note that due to the cylindrical shape of the single crystal along [$0,1,0$] nonmagnetic 
background could be reduced by narrowing the horizontal beam width.
The population of single-{\bf Q} domain was estimated to be
more than 99~\% from the intensity ratio of the magnetic satellites 
around the ($0,0,1$) and ($1,0,0$).
The final energy of TOPAN was fixed at 14.7~meV. 
Two types of horizontal collimation sequence were utilized, 
Blank($60'$)-$30'$-$60'$-Blank($100'$) and 
$30'$-$30'$-$10'$-Blank($100'$) from before the monochromator to after the analyzer.
The energy resolution of each collimation is 
evaluated to be 1.4 and 0.8 meV in FWHM, respectively. 
Higher order neutrons were removed by means of a pyrolytic graphite filter. 
Furthermore, in order to reduce the high-energy neutron background 
a sapphire single crystalline filter was inserted in 
between the first and second Soller collimators.

\begin{figure}
\centering
 \includegraphics*[scale=0.55]{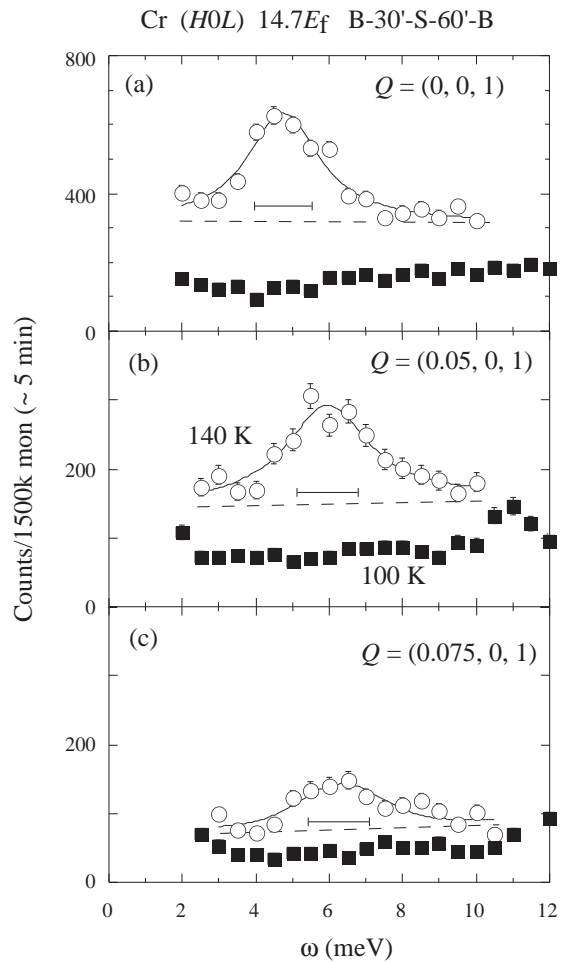} \vskip 4pt
\caption{Constant-$\bf Q$ scans along the [$\zeta,0,0$] direction around ($0,0,1$).
In the LSDW phase the scattering is small (filled squares) and independent of $\omega$.
The scattering in the TSDW phase (open circles) clearly shows a peak
that moves to higher $\omega$ with increasing $\zeta$.
Each horizontal bar below the peak represents the energy resolution.
}
\label{Fig3}
\end{figure}

\begin{figure}
\centering
 \includegraphics*[scale=0.53]{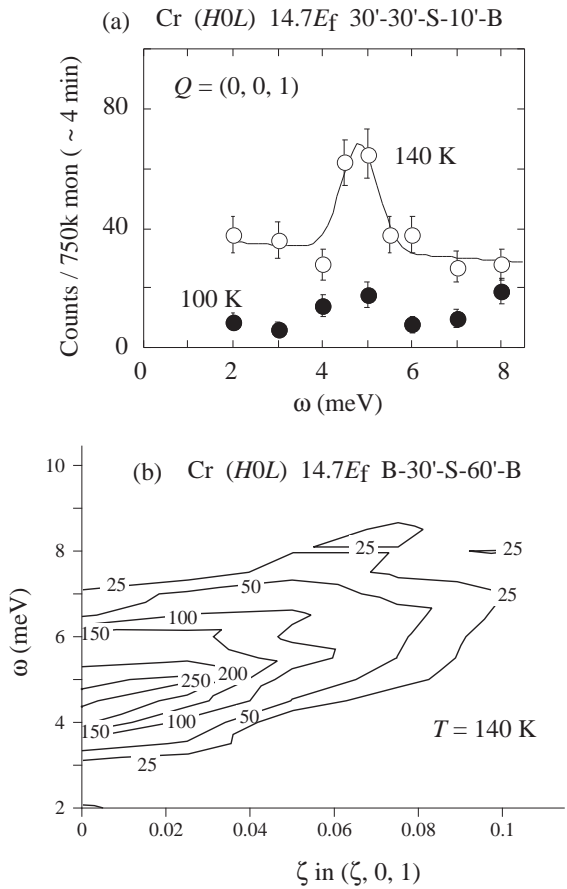} \vskip 4pt
\caption{
(a) High-resolution energy specta at ($0,0,1$).
The Fincher excitation in the TSDW state ($T = 140$ K)
shows a single peak.
(b) Contour-map of the magnetic scattering due to the
Fincher excitations, measured with the medium resolution.
It shows a gapped dispersion that extends with
increasing $\zeta$ towards 8~meV. 
The energy-independent intensities are subtracted.
}
\label{Fig4}
\end{figure}

Typical scans for scattering vectors $\bf Q$ along (scan A) 
and perpendicular (scan B) to the incommensurate
vector $\bf  Q_\pm$ are shown in Figs.~\ref{Fig2}(a) and (b), respectively.
As seen in Fig.~\ref{Fig2}(a), 
the scattering in the TSDW phase ($T = 140$ K) is composed of the
Fincher-excitation seen at around ($0,0,1$) and incommensurate peaks at $(0,0,1\pm \delta$).
In this scan there exists a 
substantial difference in the intensities between the TSDW and LSDW phases.
The broad scattering  centered at ($0,0,1$) in the LSDW phase ($T = 100$ K) is 
not due to the Fincher-mode as the intensity at ($0,0,1$) exhibits no 
appreciable energy dependence (Figs.~\ref{Fig3}(a) and \ref{Fig4}(a)). 
The Fincher mode is easily observed on top of the energy independent magnetic
intensity (broken line in Fig.~\ref{Fig3}(a)).
On the other hand, in the scan B no remarkable difference is seen between the two phases. 
The Fincher-mode is therefore not observed in this scan. 
On this aspect many detailed discussions have already been made. 
Note that in the previous scans along  $\bf Q_\pm$ the
weak signal from the FB-mode is difficult to detect 
due to the steep "background" from the incommensurate peaks. 

In order to get a clue about the origin of the FB-mode we explored the $\bf
Q$-dependence of the Fincher excitation perpendicular to $\bf
Q_\pm$ (scan (b) in the inset of Fig.~\ref{Fig1}).
Some typical scans measured at $T = 140$ K and $T = 100$ K are shown 
in Fig.~\ref{Fig3}. 
A well-defined peak is observed in the TSDW phase at $T = 140$ K.
Quite surprisingly, the peak energy clearly moves to larger $\omega$ with increasing
transverse momentum (see Fig.~\ref{Fig1}(b)) 
accompanied by a substantial decrease in the intensity.  
The peak-width in Fig.~\ref{Fig3} is broader than the instrumental resolution width.
It is noted that 
a well-defined signal was obtained due to the focusing effect of the instrumental resolution.

Because of this interesting observation we decided to map out the
magnetic scattering in more detail and performed constant-$\bf Q$
scans to construct contours. 
As shown in Fig.~\ref{Fig4}(a), 
we can get a clear excitation spectrum with high-resolution,
which confirms the single-peaked Fincher excitation.
However, we tuned the spectrometer with the medium collimator sequence
so that we could get a resonable statistics for the contours.
We note that
the signal-to-noise ratios in Figs.~\ref{Fig3}(a) and \ref{Fig4}(a)
do not change so much even in the different instrumental resolutions.
The intensity of the new mode 
as well as the Fincher-mode were evaluated by subtracting the nearly constant
intensity (broken lines in Fig.~\ref{Fig3}) as described before.
The resulting intensity contours are shown in Fig.~\ref{Fig4}(b).
One can follow the dispersion relation out to $\zeta \simeq 0.1$,
i.e. about twice as far as the incommensurability $\delta$ of the
spin-density wave.

Our results establish a new magnon branch centered around the
Fincher-mode at $(0,0,1)$ and shows a weak dispersion along the
$h$ and $k$ direction, i.e. perpendicular to the magnetic ordering
vector $\bf Q_\pm$. In contrast to previous scans along the $l$
direction the data is rather clean because there is no strong
incommensurate scattering near the "silent" positions $(\pm
\zeta,0,1)$. The new mode has the following important properties:
i) It has an energy gap of 4 meV and can only be observed in a
relatively narrow $\bf Q$ and $\omega$ range. ii) The new mode
exists only in the TSDW phase demonstrating that it is only
allowed due to the transverse orientation of the magnetic moments
with respect to $\bf Q_\pm$, i.e. the spin-flop transition opens a
new degree of freedom for excitations. iii) The mode has the same
longitudinal polarization as the Fincher-mode implying once more
that it is intimately connected with the ordering of the
spin-density-wave. It is clear these results are not compatible
with any interpretations for the FB-mode given in the literature
so far.

The non-existence of the mode in the LSDW-phase and the gap in the
TSDW-phase bear some similarity to optical phonon modes in
insulators that show different dispersions depending on their
polarization being transverse or longitudinal with respect to the
direction of propagation. One may speculate that it is possible to
excite domain walls (or stripes) in the TSDW phase that require a
nucleation energy of about 4 meV and propagate perpendicular to
$\bf Q_\pm$ thus causing a dispersion. This process may be
energetically less favorable in the LSDW-phase.

We point out that the new mode has also some intriguing
similarities to the resonance-like peaks in the dynamical magnetic
susceptibility observed for the several systems with strongly
correlated electrons; high-$T_{\rm c}$ cuprates such as
YBa$_2$Cu$_3$O$_{7-y}$~\cite{fong96} 
that orders also in an incommensurate structure, 
and the geometrically frustrated ZnCr$_2$O$_4$~\cite{lee00}.
We hope that our results
encourage new efforts to understand the antiferromagnetic state in
Cr, in particular the interplay between this new mode and the
enhancement of the longitudinal incommensurate scattering and the
scattering at the silent positions, which will also elucidate the
common aspect of magnetic excitations in strongly correlated
electron systems.

\begin{acknowledgments}


We would like to thank S. H. Lee, T. M. Rice, R. Pynn and 
J. M. Tranquada for stimulated discussions.
The present research was supported by the U.S.-Japan Cooperative Neutron-Scattering Program.
The work at Tohoku and Kyoto was supported by the Ministry of Monbu-Kagaku-shou of Japan,
and the Core Research for Evolutional Science and Technology Project.
Work at Brookhaven is supported by the Division of
Material Sciences, U.S. Department of Energy under contract DE-AC02-76CH00016.

\end{acknowledgments}


\bibliography{basename of .bib file}

\end{document}